# Gender Differences in Research Topic and Method Convergence among Collaborating Scholars in Library and Information Science

Chengzhi Zhang, Linlei Xie, Siqi Wei
Department of Information Management, Nanjing University of Science and Technology, Nanjing, 210094, China
Corresponding author: Chengzhi Zhang (zhangcz@njust.edu.cn).

**Abstract:** This study explores gender differences in research topic choice and methodology among collaborating scholars. Previous studies have often focused on gender differences in research topics or methods at the individual level of scholars, without considering collaborating groups, lacking depth and practical guidance. This study takes Library and Information Science (LIS) as an example, employing the Top2Vec method for topic identification and the CogFT model for research method classification. It systematically analyzes 25,204 papers published between 1990 and 2022 to investigate gender differences in the convergence of research topics and method choices among collaborating scholars in this field. The results of the study found that female scholars showed lower convergence in their research methods and topic choices compared to male scholars. This study uses a relatively systematic methodology to address the difficulty of studying gender differences in academic publishing, and is expected to serve as a reference for other disciplines and research questions. This study also emphasizes the manifestation of gender differences in collaborative research and provides insights into the convergence and diversity of research topics and methods chosen by scholars.

## 1. Introduction

With the continuous advancement of scientific research, academic collaboration has become a critical driver of knowledge innovation. Collaboration between scholars goes beyond joint publications, involving deep interactions in exploring knowledge, exchanging ideas, and solving problems. There is often a certain "convergence" in the selection of research topics and methods by collaborating scholars, that is, the degree of consistency in their choices. This convergence reflects the core concerns of the academic field and hints at potential trends in its development. Research suggests that male and female scholars may tend to explore different topics and use different methods (Chari & Goldsmith-Pinkham, 2017; Key & Sumner, 2019). So, what happens when they collaborate? Do we still see these trends? This question is important because understanding the difference between how men and women typically work can facilitate better communication between male and female scholars, enable them to learn from each other's strengths, and help advisors and mentors avoid gendering their advice and instruction.

Gender differences, are influenced by physiological, psychological, and environmental factors that affect various aspects of academic life, including resource access and career development (Blau & Kahn, 2017; Fox, 2005; Hyde & Mertz, 2009; Skjelsbaek, 1998). Integrating gender analysis into research frameworks has been shown to reduce bias, promote scientific discovery, and advance social equality (Tannenbaum et al., 2019). The gender of this study only includes males and females. Although in gender studies, men and women are not binary opposites. However, current research, especially in LIS (Bisaria & Jaiswal, 2019), Scientometrics (Fell & König, 2016) disciplines and other fields (Nunkoo et al., 2020; Alers et al., 2014) generally use binary classification based on gender. On the one hand, this is because binary gender classification (male/female) has long dominated social norms and legal systems, and publicly available datasets and official statistical records rarely contain non binary gender labels. Existing tools, such as Genderize.io also rely on historical data on binary name-gender associations. On the other hand, the integration of non-binary classification across time datasets faces significant methodological challenges.

Despite a relatively balanced gender ratio in LIS, female scholars are underrepresented in top academic journals, particularly those focused on information science (Lund & Shamsi, 2023). This study focuses on Library and Information Science (LIS) to explore the difference between scholars' individual research topics and methods as well as those of other scholars within their collaborating groups, and how different combinations of male and female affect this difference.

The innovation of this study is reflected in the following three aspects. Firstly, unlike previous studies that focused solely on a single dimension of research topics or methods, this study systematically examines the combined impact of gender differences in collaborating research on the convergence of research topics and methods. Secondly, based on a large-scale dataset in the LIS field (25,204 papers), this study employs quantitative analysis methods to provide more representative empirical evidence. Lastly, the study finds that female scholars exhibit greater diversity in research topics and methods in collaborating research. This not only expands the understanding of the mechanisms of academic team collaboration but also offers new perspectives for understanding gender differences in academic research. Therefore, this study not only fills the gap in research on gender convergence from the perspective of academic collaboration but also provides practical guidance for fostering a more inclusive academic environment and policy-making.

## 2. Problem statement

A lot of attention has been paid to the differences in the choice of research topics and methods among scholars of different genders. (Su & Rounds, 2015; Thelwall et al., 2019; Plowman & Smith, 2011; Williams et al., 2018). However, the universality of these findings remains to be further validated, and there is a lack of in-depth analysis of the underlying factors behind gender differences. Additionally, few studies have delved into the convergence in topic and method selection among collaborating scholars with different gender combinations. Exploring this disparity not only aids in understanding the role of gender in academic collaboration but also provides empirical support for the advancement of gender balance and diversity in academia.

Taking the LIS field as an example, this study selected research papers from 14 recognized and authoritative international core journals in this field between 1990 and 2022 as the research sample. The aim is to systematically investigate gender differences in the convergence of topic and method choices among collaborating scholars. This paper is an empirical analysis, with its research focus not on the enhancement of certain methods or models per se. While manual processing of some data in this study might yield greater precision, the sheer amount of work involved makes this approach impractical. Consequently, we employed methods previously utilized by the authors' research group to manage this large-scale dataset, striking a balance between precision and operational feasibility. This represents a quintessential data-driven research endeavor. Specifically, this study will address the following questions:

**RQ1:** In the field of Library and Information Science (LIS), are there convergence differences in topic and method selection between collaborating groups of different gender compositions?

**RQ2:** If such differences exist, what tendencies do collaborating groups composed of different genders exhibit in choosing specific topics and methods?

By answering these questions, this study is anticipated to uncover the specific manifestations of gender differences in academic collaboration, providing novel perspectives and ideas for promoting gender balance and diversity in the academic community. Additionally, the findings of this study will serve as valuable references for scholars in other fields, stimulating further thought and discussion on gender differences in academia.

# 3. Literature review

### *3. 1 Gender differences in scholars' research topic and method selection*

The difference in career choices between female and male is a significant factor contributing to women's lagging behind men in the overall scientific field. In specific academic fields, this discrepancy manifests in research topic (Ceci & Williams, 2011）. Related studies have found that in management (Nielsen & Borjeson, 2019), natural language processing (Vogel & Jurafsky, 2012),medicine (Alers et al., 2014; Nielsen et al., 2017）and fields such as economics and political science (Chari & Goldsmith-Pinkham, 2017; Key & Sumner, 2019), Men and women have different tendencies in choosing research topics. Female and male scholars in LIS prefer "usage studies" and "bibliometric analysis" respectively (Bisaria & Jaiswal, 2019). Some studies attempt to explain gender differences in research topics through a unified dimension, such as female and male scholars being more interested in topics related to people or objects, respectively (Su et al., 2009; Su & Sounds, 2015; Thelwall et al., 2019); or women are more concerned about social issues, while men are more enthusiastic about solving scientific problems (Zhang et al., 2021).

The differences in the nature of research topics often determine the choice of research methods among scholars of different genders (Creswell, 1994, 2003). Research in the field of management suggests that women are more inclined to engage in qualitative research, while men and mixed gender teams are more inclined to conduct quantitative research (Plowman & Smith, 2011; Williams et al., 2018; Diaz-Kope et al, 2019). Similar results have also been observed in the field of tourism (Nunkoo et al., 2020). Zhang et al. (2023) investigated gender differences in the use of 16 research methods in the field of LIS and found that female scholars prefer to use interviews, while male scholars prefer theoretical methods. However, existing research lacks a comprehensive consideration of research topics and methods. The specifics of the collaboration group were also not explored.

### *3. 2 The convergence of topic selection by collaborating scholars*

In recent years, some studies have emerged in the academic community, focusing on the convergence of research topic selection among collaborating scholars. The research topic not only encapsulates the core content of the academic paper, but also reflects the research direction and interests of the scholar, which are intricately linked to their academic performance. Bu et al. discovered that collaborators of high-impact researchers tend to select different research topics (Bu et al., 2018). However, Smith et al. revealed an inverted U-shaped relationship between research topic similarity and the probability of collaboration (Smith et al., 2021). Huang et al.'s research indicates that highly productive scientists tend to explore novel topics to sustain innovation (Huang et al., 2022). Another controversial study suggested that topic choice is a significant factor in the lower funding rates awarded to African American and black scientists (Hoppe et al., 2019; Lauer et al., 2021). Santos et al. explored the convergence factors that contribute to research collaboration and repeated collaboration, along with their relative importance, and discovered that the impact of gender convergence on research collaboration varies across different subject areas, with the medical field experiencing the greatest influence (Santos et al., 2024).

### *3. 3 Classification of research method*

The standardization of research methods underpins discipline maturity, especially in LIS. Järvelin & Vakkari (1993) proposed a classification system encompassing 13 research methods, which set the benchmark for constructing a research methodology system in the LIS field. Chu (2015) found that more scholars in the field of LIS tend to focus on content analysis, experimental methods, and theoretical research methods. In their subsequent research, Chu & Ke (2017) classified research method in the LIS field into 16 categories, focusing on data collection. Recently, in international research on research method in the LIS field, the methodological systems utilized are primarily based on Järvelin & Vakkari's framework, Chu's methodology, or a combination of both, such as in the works of Tuomaala et al. (2014), Zhang et al. (2021), Lund & Wang (2021), Lou et al. (2021), and Zhang et al. (2023).

Quantitative research in LIS methods faces limitations in manual annotation, which has sparked people's interest in automation. Eckle Kohler et al. (2013) achieved 0.68 F1 through multi label classification of abstracts and also advocated the use of full-text content. Zhang (2021) used rule-based automatic classification and later improved it with full-text semi automation (Zhang & Zhang, 2021). Zhang (2023) significantly enhanced performance, achieving an F1 score of 0.85 through text summarization. This study further investigates gender differences in methodological convergence between topics and collaborators. Analyzing comprehensive scholar data, it calculates probability distribution disparities, enhancing accuracy and interpretability.

## 4. Data and methodology

In this study, we use literature data from 14 core LIS journals. We mainly use Top2Vec to identify the topic of the article, uses CogFT for method classification, and combine multiple methods such as gender identification API tools to identify the gender of scholars. In addition, we use JS divergence to measure the convergence of theme and method selection among different gender collaboration groups.

### *4. 1 Data Sources*

This study mainly focuses on LIS scholars, using data from 14 core LIS journals selected from the 2022 Journal Citation Reports (JCR) - LIS Q1-Q3 sections and Järvelin & Vakkari (2022) catalog. It covers data from 25204 research papers between 1990 and 2022, excluding editorials and book reviews. It mainly includes two parts: full-text data for topic and method analysis, and metadata for scholar disambiguation and gender identity. This broad span captures the repositioning of themes and methods in the field over the past thirty years (Järvelin & Vakkari, 2022; Ma & Lund，2021).The number of research papers in each journal is shown in Table 1, with CIM contributing the most, accounting for over 20% of publications, followed by JASIST and IPM, which together account for nearly half of the dataset.

**Table 1.**List of source journals and the number of research papers.

| Journal Name (abbreviated) | # Articles |
|---|---|
| Aslib Journal of Information Management (AJIM) | 1,300 |
| College and Research Libraries (CRL) | 1,282 |
| Information Processing and Management (IPM) | 2,710 |
| Information Technology and Libraries (ITL) | 530 |
| International Journal of Information Management (IJIM) | 1,823 |
| Journal of Documentation (JOD) | 1,353 |
| Journal of Information Science (JIS) | 1,406 |
| Journal of Librarianship and Information Science (JLIS) | 799 |
| Journal of the Association for Information Science and Technology (JASIST) | 3,837 |
| Library and Information Science Research (LISR) | 749 |
| Library Quarterly (LQ) | 452 |
| Online Information Review (OIR) | 1,569 |
| Scientometrics (SCIM) | 5,611 |
| The Electronic Library (TEL) | 1,783 |
| Total | 25,204 |

### *4. 2 Research topic identification*

This study used the Top2Vec[1] method to model the topics of 14 foreign journal articles using their titles and abstracts. We firstly use NLTK (Natural Language Toolkit)[2] to clean the data of titles and abstracts. The topic model we employ, Top2vec, initially maps both documents and words into the same semantic vector space using Doc2vec. Subsequently, it utilizes the density-based clustering algorithm HDBSCAN to perform clustering, where each cluster obtained represents a distinct topic. Finally, the words corresponding to the N nearest word vectors to the topic vector are selected to represent the topic. The topics generated using the Top2Vec method are more informative than PLSA and LDA (Angelov, 2020). PLSA (Probabilistic Latent Semantic Analysis) uses a generative latent class model to perform a probabilistic mixture decomposition to obtain topics (Hofmann, 2001). LDA (Linear Discriminant Analysis) infers topics through document word matrices and Dirichlet priors (Blei et al., 2003). In addition, although the clustering process of the BERTopic model is similar to the Top2Vec model, it often generates more outliers in the topic allocation process, which is not suitable for the data in this study (Grootendorst, 2023).

[1] https://top2vec.readthedocs.io/en/stable/Top2Vec.html
[2] https://www.nltk.org

The study used context-based coherence values (CV) to evaluate topic modeling and found that 22 topics were optimal. We use the top 10 most relevant topic words output by the Top2Vec model for each topic to name the topics. When examining the topic keywords generated by the Top2Vec model, we found that two out of 22 topics had semantic overlap with the other two topics (Appendix A). Referring to the theme classification system developed by Järvelin & Vakkari, merge these two topics with the other two and summarize them using more common topic names (Järvelin & Vakkari, 2022). The results of theme recognition and naming are shown in Table 2.

**Table 2.** Identified topic names.

| Topic | Number of articles |
|---|---|
| Bibliometrics /Scientific evaluation | 2093 |
| Enterprise information management | 1762 |
| Information retrieval models and evaluation | 1673 |
| Information storage/Classification and indexing | 1636 |
| Scholarly communication /International collaboration | 1549 |
| Text processing/Retrieval | 1531 |
| Information ethics/Philosophy of information | 1497 |
| Information behavior/User study | 1450 |
| Information communication | 1418 |
| Information organization/Information management | 1271 |
| Academic publishing/Open access | 1204 |
| Information seeking behavior | 1160 |
| Information literacy teaching/cultivation | 1102 |
| Webometrics | 1021 |
| Medical/Health information services | 1014 |
| LIS career research | 883 |
| Social media/ Public opinion analysis | 791 |
| Scientometrics/Informetrics | 742 |
| Digital cultural heritage management | 716 |
| Intellectual property/Innovation management | 692 |

Topic 1, bibliometrics / scientific evaluation, emerged from the combination of two topics and accounted for the largest number of articles. This is logical given that the SCIM journal, which contributed the most articles to this study's dataset, specializes in this field.

### *4. 3 Research method classification*

Chu & Ke established a classification system for research method in the LIS field. This system encompasses 16 research methods, such as bibliometrics, content analysis, Delphi method and others (Chu & Ke, 2017). This study employs the CogFT (Cognitive Full-Text) model, which is an automatic classification model for research methods based on full-text cognition. It applies the CogLTX (Cognize Long TeXts) model (Ding et al.,2020) to the task of research method classification. The model has been trained on the classification system and annotated data provided by Chu & Ke to classify the research methods used in 14 LIS journal articles (Zhang & Tian, 2023). The strength of this model lies in its capability to extract key descriptions of research method from the full text, thereby classifying the research method adopted in the articles accurately.

The construction of the CogFT model is primarily divided into two stages: the full-text abstraction stage and the research method classification stage. In the full-text abstraction stage, the textual content of each document D is segmented into 128 blocks. If the text length exceeds 8192 words, the model truncates the text from the beginning. The abstraction network then feeds these input texts into SciBERT for text embedding (Beltagy et al., 2019). A global encoder based on Self-Attention is subsequently employed to capture the relationships between each block. Ultimately, via the output layer, the abstract network determines the probability of each block being a key block that describes the research method. During the research method classification stage, the four blocks with the highest probabilities and the first four blocks of the document are combined and input into the classification network. The classification network directly utilizes the BERT model (Devlin et al., 2019). The features extracted from BERT are then fed into a Sigmoid layer at the end of the network for classification. Experiments reveal that the CogFT model outperforms previous research method classification models, achieving an $F_1$ score of over 85%. Additionally, in a random sample of 200 articles with unknown method categories, the manual annotations and the model's automatic classification results exhibit a high Kappa consistency of approximately 0.81. Therefore, this study employs the CogFT model to predict the method categories used in research articles from 14 LIS journals.

Appendix B summarizes the classification results of 16 research methods, indicating the frequency of each method's usage across all research papers published in the 14 LIS journals. It is noteworthy that the total frequency of research method (28,791) exceeds the total number of articles collected in this study (25,204). This is attributed to the fact that an article may report the use of multiple research method within its research design, and the CogFT classification model employed in this study is a multi-label classification model capable of predicting multiple method utilized in a single article.

Unlike the distribution of research method in JASIST, JOD, and LISR journals presented by Zhang (2023) and Chu & Ke (2017), the most frequently used method in this study is bibliometrics, followed by experimentation, which topped their respective statistics. This difference is attributed to the selection of data sources, as the 14 journals chosen for this study contain a significant proportion of articles from SCIM, which focuses predominantly on bibliometric research. Nonetheless, the six most commonly utilized method in the LIS field remain unchanged: bibliometrics, experiment, questionnaire, theoretical approach, content analysis, and interview, consistent with previous studies. Except for the category of "other methods," the remaining method are still used less frequently due to their unique characteristics. With the expansion of the dataset in this study, the frequency of "other methods" has increased significantly, necessitating further classification in subsequent research.

### *4. 4 Scholar name disambiguation and gender identification*

Due to the prevalence of scholars with identical names in academia, the lack of disambiguation may lead to confusion in research method and topics among different scholars, thereby compromising the accuracy and reliability of research. Consequently, our initial priority lies in disambiguating scholars' names, followed by the identification of their gender.

#### *4.4.1 Scholar name disambiguation based on paper metadata*

This study selects OpenAlex[1], a large index database of academic resources, which provides disambiguation of authors, papers, journals, institutions and other related academic entity data. OpenAlex inherits most of the data from the Microsoft Academic Graph (MAG)[2] developed by free Microsoft Research and is updated every two weeks. MAG, which is widely used because it integrates information from academic publications across multiple subject areas around the world, ceased operations in 2021. Compared to MAG, OpenAlex has wider coverage and better usability, and some studies have shown that it is more suitable for bibliometric research than MAG (Scheidsteger & Haunschild, 2023).

First, this study extracted the DOI field from the metadata of academic papers. Then, utilizing the PyAlex third-party library, this study searched for detailed information on each paper based on its DOI in the OpenAlex database, obtaining author information and their corresponding unique identification IDs. This process initially achieved author disambiguation. Our study compared the number and order of authors in each paper in OpenAlex with the data collected from the website to ensure consistency. This study discovered that some authors who had only published a single paper were not included in the OpenAlex database and were absent from the author information of the corresponding article, leading to discrepancies in the number of authors. For these authors, after manual verification, this study also assigned them a manually specified unique identification ID. Through these steps, our study assigned a unique ID to each author of every paper, covering a total of 35,164 scholars.

### *4.4.2 Gender identification of scholars based on paper metadata*

This study mainly refers to the method of Zhang et al. (2023), using publicly available gender datasets and using their first names (excluding surnames) to identify the gender of scholars. Although the method proposed by Zhang et al. was relatively accurate, the recall rate was relatively low. As an improvement, this study combines it with online gender recognition API tools and uses manual search methods for scholars who cannot infer gender through these two methods. This process includes three steps:

(1) Gender recognition based on gender dataset: Using the same dataset as Zhang et al. (2023), including USSA birth name and gender data (popular baby names, 2022), PUBMED author name and gender list (Torvik & Smalheiser, 2009), and WGND (Raffo & Lax Martinez, 2018). This study used Vogel & Jurafsky's manually annotated ACL selection list for validation, with a matching accuracy of approximately 98%, inferring the gender of 26319 scholars (74.85%).

(2) Gender identification based on online API tool: This study uses two online gender identification API tools, Genni 2.0[3] and Genderize.io[4]. For this study, a prediction probability higher than 85% was set as a valid identification (Akyil et al.,2020). Another condition of Genderize.io is to ensure the accuracy of at least 10 samples in the database. Through API identification, the genders of 3,786 scholars were inferred.

(3) Gender identification method based on manual retrieval: For scholars who cannot identify gender through datasets and API tools, manual retrieval methods are used, mainly targeting neutral gender names, abbreviations, and East Asian names. Search for personal profile images, gender clues, and author introductions on search engines and academic websites to maximize recall rates. In addition, this study has not yet found a suitable way to handle authors who engage in identity recognition outside of the male/female binary structure, which is also a limitation.

[1] https://openalex. org/
[2] https://www. microsoft. com/en-us/research/project/microsoft-academic-graph/
[3] https://www. geany. org/
[4] https://genderize. io/

(4) Gender identification results: Integrating these methods, this study identified genders of 32,335 scholars (91.95%), with 12,330 (38.13%) female and 20,005 (61.87%) male. The remaining unidentified scholars accounted for 8.05%, unlikely to affect final analysis results (Larivière et al., 2013). The male-to-female ratio of scholars in 14 LIS journals is approximately 6:4, aligning with previous studies. Figure. 1 illustrates the gender ratio of scholars in 14 top LIS journals, arranged in descending order of the number of publications per journal. Only 4 journals had a female scholar ratio exceeding 50%, all with relatively few publications, suggesting male scholars have an advantage in academic output in top LIS journals.

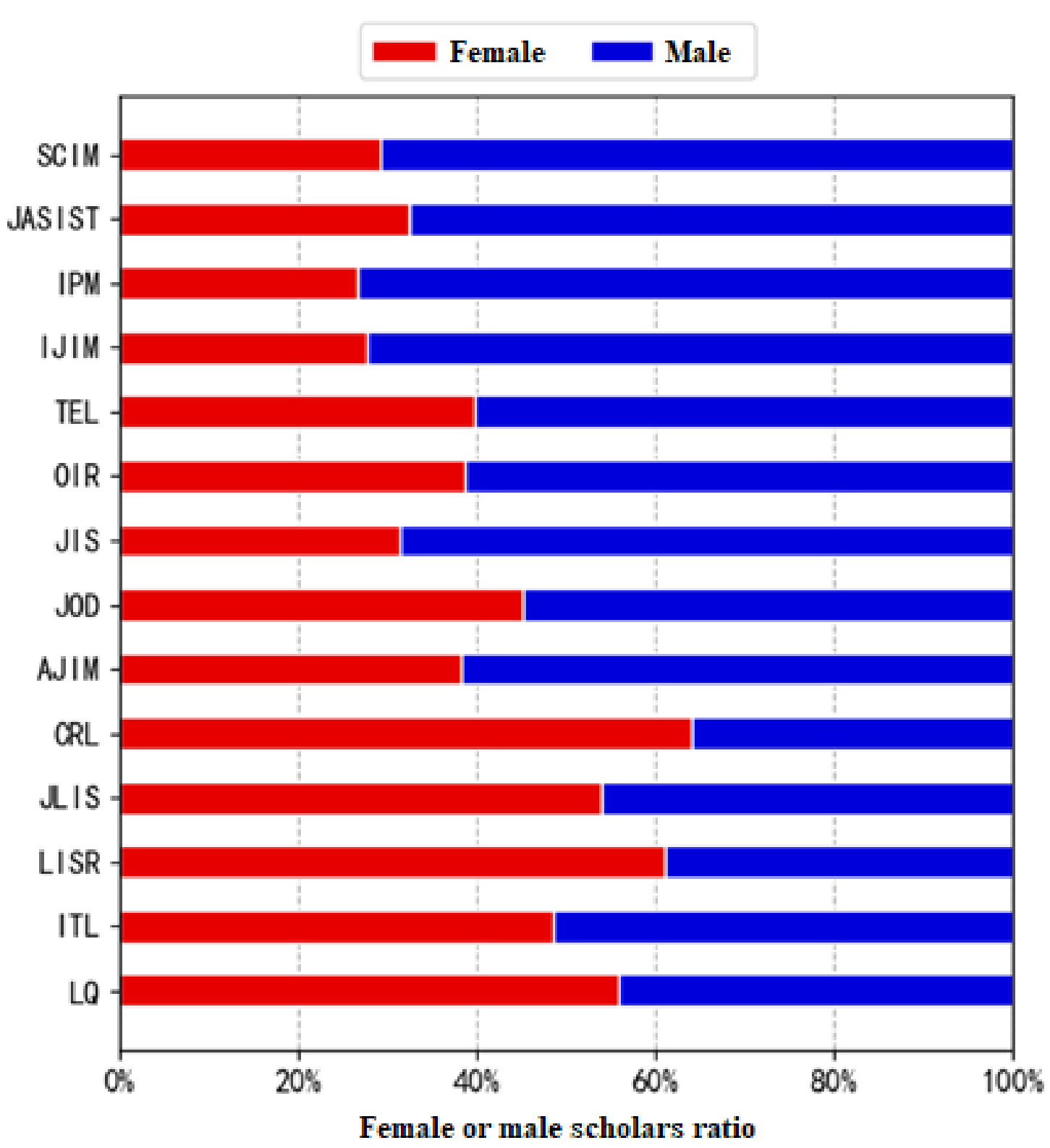

**Figure.1.** Scholar sex ratio in 14 top LIS journals.

### *4. 5 Measurement of convergence in collaborating groups of different genders*

JS divergence, also known as Jensen-Shannon Divergence, JS distance, or information radius, is a technique that assesses the similarity of information values based on their density distributions (Manning, 1999). It serves to quantify the degree of similarity between two probability distributions. Consequently, this study utilizes JS divergence to gauge the similarity between the probability distributions of topic and method selections made by two collaborating scholars.

# 5. Results

This study categorizes collaborating scholars into three types based on gender: female-female, female-male, and male-male. The objective is to analyze whether there are specific preferences or patterns in the choice of research topic and method among different gender combinations. This classification approach enables us to identify potential gender differences in the convergence of topic and method selection among collaborating scholars of varying gender combinations, explore the potential impact of gender factors in collaborating research, and how this impact shapes the direction and outcomes of such research.

## *5. 1 Gender differences in the convergence of topic and method selection*

This section addresses RQ1. The study assesses the research preferences of each scholar based on the frequency of their choice of different research topics or methods. Each scholar has a tally list for research topic and method. Given that there are 20 research topics involved in this study, the length of each scholar's topic tally list is 20. If a scholar selects Topic 1 once and Topic 4 twice in their prior research, this study increments the first element by one and the fourth element by two in their topic tally list, and so forth. The same logic applies to the scholar's method tally list. This approach enables us to derive the frequency distribution of different topics and method chosen by each scholar. By dividing these two frequency distribution lists by the number of papers published by the scholar, this study can obtain the probability distribution lists for the different topics and method preferred by each scholar.

This study calculates and compares the average JS divergence values of the topic and method selection probability distributions among female-female, female-male, and male-male collaborating scholar pairs, as presented in Table 3.

**Table 3.** Statistics of the difference of the probability distribution of topic and method selection by different collaborating Scholars.

| Collaborating scholars' choice of topics | Probability distribution mean distance for topic selection (JS divergence) | Probability distribution mean distance for method selection (JS divergence) |
|---|---|---|
| Female-Female | 0.4168 | 0.2963 |
| Female-Male | 0.4089 | 0.2938 |
| Male - Male | 0.3715 | 0.2617 |
| Overall | 0.3952 | 0.2787 |

It is evident that female scholars tend to exhibit a relatively high degree of diversity and difference in their choice of collaboration partners with regards to research topic and method. In other words, they are more inclined to collaborate with scholars who employ diverse research topic and method. In contrast, male scholars tend to prioritize the convergence of their research objects when selecting collaboration partners. However, this convergence may also constrain their research horizons and innovativeness.

Furthermore, the average JS divergence values in topic and method selection between female-male collaborating scholar pairs are similar to those between female-female collaborating scholar pairs, albeit slightly lower. This suggests that while female scholars consider the differences and complementarity of their male collaborators, they may be influenced by male scholars' preference for convergence, resulting in a slight reduction in the diversity and difference in their collaborations. This also underscores the importance of balancing the research tendencies and collaboration needs of scholars of different genders when promoting interdisciplinary and cross-gender collaboration.

### *5. 2 Gender differences in topic and method selection preferences*

This section addresses RQ2. The paper further compares the average probabilities of utilizing different research topics and methods among three distinct types of collaborating scholar pairs (female-female, female-male, male-male) with the average probabilities among all collaborating scholar pairs, as depicted in Figures 2 and 3. The horizontal axis labels represent the categories of topics and method, while the vertical axis measures the deviation between the average probability of a specific topic or method being selected by a certain type of collaborating scholar pair and the average probability of that topic or method being selected by all collaborating scholar pairs. Consequently, a value greater than 0 signifies a higher probability than average among all collaborating scholar pairs, while a value less than 0 indicates a lower probability than average.

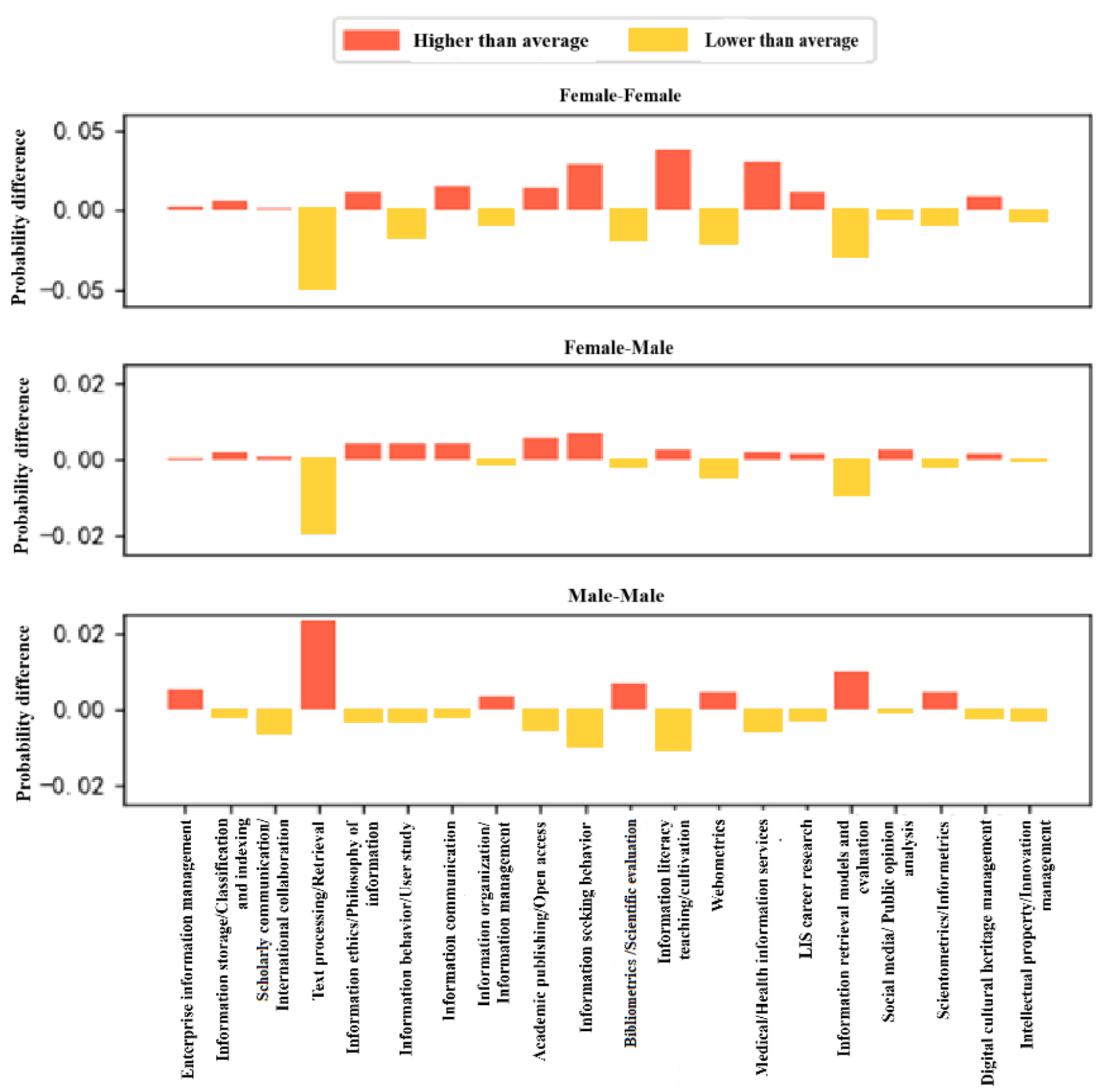

**Figure.2.** Topic selection preferences of different pairs of collaborating scholars.

Figure. 2 illustrates the disparities in topic preferences among various scholar pairs. We can see that female-female collaborating pairing shows a higher degree of diversity and balance in topic selection. This trend suggests that when female scholars collaborate with each other, they are more prone to exploring diverse fields and topics. In contrast, male-male collaborating pairs tend to have a more concentrated preference for specific topics, often focusing on in-depth research within a single domain.

Regarding topic preferences, Figure. 2 uncovers several noteworthy phenomena. Firstly, topics such as information literacy teaching/cultivation, medical/health information services, and information seeking behavior are favored by female-female collaborating pairs. This can be attributed to female scholars' heightened attention to education and health fields, as well as their distinctive perspectives on information search and processing. Moreover, the topic of text processing/retrieval is the least likely to be chosen by female-female collaborating pairs. For female-male collaborating pairs, Figure. 2 does not reveal a clear preference for any specific topic. However, they also tend to avoid the topic of text processing/retrieval. In contrast, male-male collaborators showed a greater preference for the topic, probably due to the fact that the topic involved more technical elements. Male-male collaborators are less interested in topics such as information seeking behavior and information literacy teaching/cultivation, which are about educational or user research.

It is noteworthy that the vertical axis scale of the first subfigure in Figure. 2 is relatively large, making the fluctuations in the bar charts more prominent. This underscores the diversity and depth of topic preferences among female-female collaborating pairs. They encompass numerous distinct topic areas and exhibit a relatively high level of preference for each. The depth of these preferences may reflect the in-depth research and interests of female scholars in various topics during collaborations.

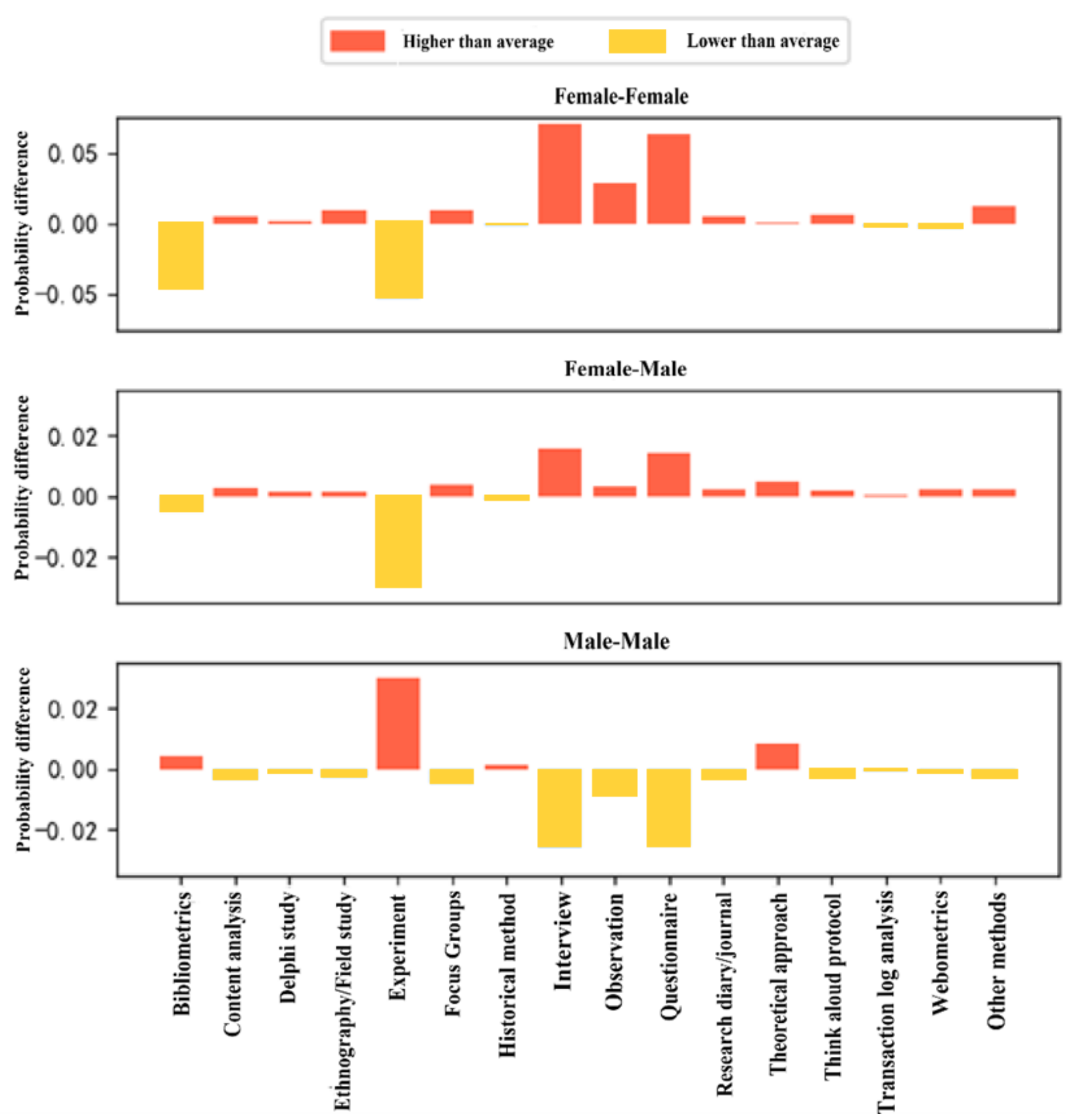

**Figure.3.** Method selection preferences of different pairs of collaborating scholars.

As illustrated in Figure. 3, female-female collaborations exhibit a greater degree of diversity and balance in the selection of research method, once again indicating that female scholars tend to have less convergence in method selection compared to male scholars. They are more inclined to opt for interview, questionnaire, and observation, which can furnish more in-depth and comprehensive research data. When female scholars collaborate with male scholars, their preference for research method resembles that of female-female collaborations, with both tending to shy away from experiment. In contrast, male-male collaborations favor experiment, which align with the precision and reproducibility of such method and may also be a characteristic tendency among male scholars in scientific research. At the same time, they also choose theoretical approach, reflecting male scholars' proficiency in constructing and validating theoretical models.

The variation in the scale values of the vertical axis demonstrates that female-female collaborations exhibit more distinct preferences in selecting research method, thereby further emphasizing the tendency of female scholars to choose complementary collaborators. They may be more inclined to collaborate with scholars who employ different research method and perspectives, aiming to enrich the breadth and depth of their research.

# 6. Discussion

## *6. 1 Implications*

### *6.1.1 Theoretical inspiration*

This study analyzed the differences in research methods and topic selection among researchers in the LISR field and innovatively used gender convergence as an analytical dimension, revealing the potential gender factors behind these differences. This discovery provides a new perspective for understanding gender imbalance in academia. In the future, the academic community can further analyze the root causes of preferences in academic cooperation, explore effective strategies to break down gender barriers, and jointly build a more just and inclusive academic environment.

This study employed a relatively systematic methodology. We combined multiple methods to improve the accuracy of gender recognition, while using deep learning models instead of traditional content analysis methods to obtain a larger corpus of research topics and methods. However, the classification performance of the research methods in this study still needs to be further improved. In the future, more advanced methods such as Large Language Model will be explored, as well as the integration of article information such as article structure. We have only experimented with literature in the LIS field, but we have adopted mainstream models for topic identification and classification of research methods, and we can also try to apply them to related research in other fields.

### *6.1.2 Practical inspiration*

The results of this study indicate that collaborating scholars are not solely determined by their own research questions when choosing research methods and topics, but are influenced by gender and its convergence differences. This difference reveals the gender imbalance and diversity present in academic research. It is necessary to understand the different preferences of scholars of different genders in academic research and team collaboration, establish a fairer and more just academic evaluation system, and ensure that female scholars enjoy the same rights and opportunities as male scholars in academic research. The academic community should have a correct understanding of the characteristics and differences between men and women in academic research, encourage scholars to participate in more interdisciplinary collaborations, and promote exchanges and cooperation in different fields. It should also provide more support and opportunities to encourage female scholars to demonstrate confidence and fully unleash their potential and abilities in research collaborations.

### *6. 2 Analysis of potential reasons for gender differences*

The investigation into gender differences in the convergence of topic and method preferences among collaborating scholars revealed that female scholars exhibit lower convergence in their topic and method preferences compared to their collaborators, regardless of whether it is a female-female or female-male combination. The similarity in topic and method preferences for these combinations is significantly lower than that of male-male combinations. This finding aligns with previous studies, such as Smith et al. (2021), who found that women tend to collaborate with colleagues who are further away in terms of research topics. The current study also uncovered that women prefer to collaborate with scholars who employ different research methods. Additional research has shown that women are more likely to engage in interdisciplinary collaborations compared to men (Lungeanu et al., 2014; Van Rijnsoever & Hessels, 2011). A potentially related finding is that the average number of co-authors in teams with female leaders (2.78) is higher than that of teams with male leaders (2.64).

The study seems to indicate that female scholars prefer to collaborate with scholars who have different research topics and methods or to participate in larger teams with a broader range of methods. In contrast, male scholars may prefer to collaborate with scholars who have similar research directions and methodologies, tending to favor smaller yet more specialized teams. Some studies have pointed out that women tend to have a more cautious and thorough work and research style compared to men, especially in STEM fields, due to the long-term underestimation of their contributions, which requires them to provide additional evidence to prove their capabilities and methods. Other research has shown that female scholars exhibit less confidence in their research, even when their professional level is comparable to that of male scholars. These factors may contribute to their tendency to conduct more comprehensive research and choose more complementary collaborators, though this is just one possible explanation (Fehr, 2011; Leuschner, 2015; Liberatore & Wagner, 2022; Morris, 2015).

### *6. 3 Limitations and future work*

The study aims to explore the gender differences in the convergence of topic and method preferences among collaborating scholars in a specific field. However, given the intricate nature of factors contributing to gender disparities and inequality in academic domains, this study falls short of considering all of them, and thus has some limitations. Firstly, the research data source is confined to international core journals within a specific field, excluding articles authored by scholars in non-field-specific journals. Due to the long time span covered by the data in this study, there may be significant differences between the current academic understanding and discourse on gender and those of 1990. The statistical analysis results of this study may not adequately highlight the differences across different years. Secondly, the study primarily relies on quantitative analysis to examine gender differences in the convergence of topic and method preferences, which may inherently lead to some gender biases embedded in quantitative methodologies. In addition, this study has limitations in determining the authors' gender based on their names. Furthermore, the study neglects factors such as journal review bias, which could potentially explain the gender disparities in topic and method preferences among scholars. Lastly, as the research topics and methods extracted from academic paper texts are categorical data, it becomes challenging to quantify their impact on other continuous variables, such as the number of publications and citations.

Future research in this area could include: expanding the dataset to include articles published by LIS scholars in non-field-specific journals or restricting the data source to scholars within specific subfields. It could also further explore gender differences in theme and method selection among collaborating groups in different periods based on years. In the future, we will also explore ways to enhance the performance of research method classification, gender recognition, and other methodologies to achieve more accurate results. We will conduct qualitative research to further validate the findings through interviews with female and male scholars; incorporate additional information about scholars to aid in determining their gender, beyond just their names; consider potential journal review biases; and explore the impact of gender differences on the convergence of research topic and method preferences among scholars by changing variable types or constructing new measurement indicators.

## 7. Conclusion

This study delved into the gender differences in topic and method preference convergence among collaborating scholars in the field of LIS and found significant results. The survey shows that the convergence of theme and method preferences between female scholars and their collaborators is significantly lower than that of male-male pairings, whether in female-female pairing or female-male pairing. By introducing a new perspective on gender issues, this study not only supplements the theoretical foundation for building a more inclusive and open academic environment, but also further inspires profound thinking on how to promote cooperation between scholars of different genders in practical operations. These findings are expected to translate into specific policy recommendations to promote gender balance and diversity in academia, thereby fostering the flourishing development of academic cooperation and widely promoting the achievement of gender equality as a social goal.

### Acknowledgments

This work was supported by the National Natural Science Foundation of China (Grant No.72074113).

## Appendix

### Appendix A. Details in Research topics

| No. | Topic name | | Keywords (Top10) | # articles |
|---|---|---|---|---|
| 1 | Bibliometrics /Scientific evaluation | Bibliometrics | laureates, referee, laureate, hirsch, productivity, prestigious, nobel, scientist, fellowship, award | 2,092 |
| | | Science evaluation | JIF(Journal Impact Factors), hirsch, percentile, citation, counting, egghe, fractional, JCR, indicator, citedness | |
| 2 | Enterprise information management | | agility, organisational, enterprise, EDI(Electronic Data Interchange), BDA(Big Data Analytics), cybersecurity, outsource, SMEs(Small and Medium-sized Enterprises), corporate, strategic | 1,762 |
| 3 | Information retrieval Models and evaluation | Information retrieval evaluation | TREC(Text REtrieval Conference), retrieval, LSI, assessor, IR(Information Retrieval), relevance, topicality, judgment, CLIR(Cross-Language Information Retrieval), boolean | 1,673 |
| | | Information retrieval model | CLIR, monolingual, corpus, translation, morphological, language, dictionary, stemmer, arabic, bilingual | |
| 4 | Information storage/Classification and indexing | | cdrom, workstation, OPAC(Online Public Access Catalogue), catalogue, electronic, nigerian, disc, OPACs, desktop, gateway | 1,636 |
| 5 | Scholarly communication/International collaboration | | cuban, country, biotechnology, mexico, sci, internationally, brazil, collaboration, international, France | 1,549 |
| 6 | Text processing/Retrieval | | LSI(Latent Semantic Indexing), extractive, compression, CBIR(Content-Based Image Retrieval), compress, summarisation, summarization, keyphrase, string, unlabeled | 1,531 |
| 7 | Information ethics/Philosophy of information | | epistemology, epistemological, philosophical, librarianship, discursive, critique, rhetorical, discourse, conception, serendipity | 1,497 |
| 8 | Information behavior/User study | | intention, perceive, antecedent, consumer, loyalty, perceived, continuance, hedonic, enjoyment, shopping | 1,450 |
| 9 | Information communication | | roundup, censorship, freedom, democracy, ALA(American Library Association), right, cdrom, society, entitle, eugene | 1,418 |
| 10 | Information organization/Information management | | hypertext, ontology, metadata, thesauri, xml, faceted, LCSH(Library of Congress Subject Headings), thesaurus, folksonomies, RDF(Resource Description Framework) | 1,271 |
| 11 | Academic publishing/Open | | journal, citation, scholarly, scopus, WOS, | 1,204 |

| | | | |
|---|---|---|---|
| | access | arxiv, altmetrics, cite, OA(Open Access), publish | |
| 12 | Information seeking behavior | engine, search, searcher, searching, session, usability, log, OPAC, navigational, interface | 1,160 |
| 13 | Information literacy teaching/cultivation | student, instruction, classroom, literacy, instructional, teach, pedagogical, course, skill, IL(Information Literacy) | 1,102 |
| 14 | Webometrics | cocitation, coupling, visualize, ACA(Author Cocitation Analysis), map, visualization, scaling, mapping, specialty, intellectual | 1,021 |
| 15 | Medical/Health information services | mother, adult, immigrant, illness, patient, refugee, health, youth, everyday, teen | 1,014 |
| 16 | LIS career research | employer, job, ARL(Association of Research Libraries), salary, librarian, qualification, librarianship, duty, profession, tenure | 883 |
| 17 | Social media/ Public opinion analysis | twitter, tweet, weibo, emergency, post, microblogs, disaster, misinformation, medium, microblog | 791 |
| 18 | Scientometrics/Informetrics | lotka, bradford, Egghe(Leo Egghe), mathematical, exponent, zipf, informetric, obsolescence, exponential, poisson | 742 |
| 19 | Digital cultural heritage management | preservation, archive, digitization, heritage, digital, archival, archivist, digitisation, museum, digitize | 716 |
| 20 | Intellectual property/Innovation management | patent, uspto, EPO(European Patent Office), trademark, assignee, patenting, technological, biotechnology, invention, nanotechnology | 692 |

**Appendix B.** Results of research method classification

| No. | Research method | # articles | Percentage(%) |
|---|---|---|---|
| 1 | Bibliometrics | 5754 | 19. 99 |
| 2 | Experiment | 4866 | 16. 90 |
| 3 | Questionnaire | 4135 | 14. 36 |
| 4 | Theoretical approach | 4124 | 14. 32 |
| 5 | Content analysis | 3934 | 13. 66 |
| 6 | Interview | 1872 | 6. 50 |
| 7 | Other method | 930 | 3. 23 |
| 8 | Historical method | 761 | 2. 64 |
| 9 | Webometrics | 693 | 2. 41 |
| 10 | Transaction log analysis | 526 | 1. 83 |
| 11 | Observation | 372 | 1. 29 |
| 12 | Focus Groups | 250 | 0. 87 |
| 13 | Ethnography/Field study | 230 | 0. 80 |
| 14 | Think aloud protocol | 175 | 0. 61 |
| 15 | Research diary/journal | 95 | 0. 33 |
| 16 | Delphi study | 74 | 0. 26 |
| | Total | 28791 | 100 |